\documentclass[a4paper,showkeys,floatfix,aps,pre,reprint,groupedaddress]{revtex4-1}
\usepackage{graphicx}
\usepackage{amsmath, amssymb, amsfonts}
\usepackage{color}
\usepackage[caption=false]{subfig}

\begin{document}

\title{Self-organized dynamics in local load sharing fiber bundle models}



\author{Soumyajyoti Biswas}
\email[]{soumyajyoti.biswas@saha.ac.in}
\author{Bikas K. Chakrabarti}
\email[]{bikask.chakrabarti@saha.ac.in}

\affiliation{
Condensed Matter Physics Division, Saha Institute of Nuclear 
Physics, 1/AF Bidhannagar, Kolkata-700064, India.\\
}

\date{\today}

\begin{abstract}
\noindent  We study the dynamics of a local load sharing fiber bundle model in two dimensions, 
under an external load (which increases
with time at a fixed slow rate)  applied at a single point. Due to the local load sharing nature, the redistributed load 
remains  localized along
the boundary of the broken patch. The system then goes to a self-organized state with a stationary average value of 
load per fiber along the (increasing) boundary of the broken patch (damaged region) and a scale free distribution of  
avalanche sizes and other related 
quantities are observed.  In particular, when the load redistribution is only among nearest surviving fiber(s), 
the numerical estimates of the exponent values are comparable with those of the Manna model. When the load 
redistribution is uniform
along the patch boundary, the model shows a simple mean-field limit of this self-organizing critical behaviour, 
for which we  give analytical estimates of 
the saturation load per fiber values and avalanche size distribution exponent. These are in good agreement with 
numerical simulation results. 
\end{abstract}

\maketitle

\section{Introduction} 
\noindent The fiber bundle model, since its introduction \cite{first}, has been studied widely as a prototypical model of failure dynamics 
\cite{dani,coleman,books, rmp1}. 
This discrete element model, involving disorder and non-linear dynamics (due to thresholds), also enables engineers to apply it to analyse the 
breaking properties
of real materials (e.g,, fiber reinforced composites \cite{kun}) and its simplicity and occasional analytic tractability \cite{rmp1}
has attracted statistical physicists, specifically for its intriguing dynamical critical behaviors  \cite{newman,roux, pradhan1,hemmer1,hansen1,sornette,zapperi,pradhan2,pradhan2.5,pradhan3,pradhan4}.

There are mainly two extreme versions of this model. Both the versions consider a bunch of fibers (Hooke springs) hanging from a rigid ceiling
and a platform is connected to the ends of these fibers and a load hangs from that platform. 
Each fiber has a given limit or load-carrying threshold
 (usually taken randomly from some distribution function), beyond which it fails. Completely different behaviors are 
observed when the elastic property of the lower platform changes. Two extreme cases arise when the platform is either absolutely
rigid (global load sharing case) or absolutely soft (though inextensible; local load sharing case).
 When a fiber breaks in the former case, 
its extra load is equally shared by all other remaining fibers, due to rigidity of the lower platform. 
In the latter case, however, due to local deformation of the platform, the load
of the broken fiber is only to be carried by nearest surviving neighbours (stress concentration occurs around the failure 
or breaking). 
\begin{figure}[tbh]
 \includegraphics[width=5cm]{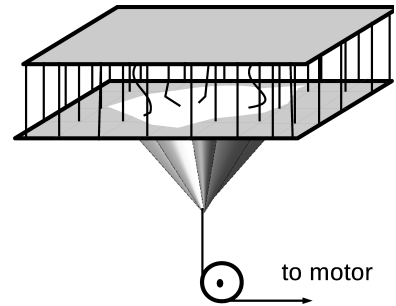}
   \caption{A schematic diagram of our model. Load is slowly increased at a single point on the 
lower platform. The patch of broken fibers is indicated by the white portion in the lower platform.}
\label{setup}
\end{figure}
While in the global load sharing version the load at
which the system completely fails scales with system size linearly, 
for the local load sharing case  the increase is only sub-linear
(in fact $N/log(N)$; $N$ being system size \cite{rmp1}). 
The implication being, the critical load per fiber $\sigma_c$ at which the
system fails, becomes finite for the global load sharing case and  goes to zero 
in the large system size limit for the local load sharing case. 
Therefore the observations like divergence of relaxation time, proper scale free size
distribution of avalanches, if any, are not seen in the local load sharing case;
 unlike in the global load sharing version, where these can be analyzed in detail \cite{rmp1}. 

\begin{figure}[tbh]
 \includegraphics[width=7cm]{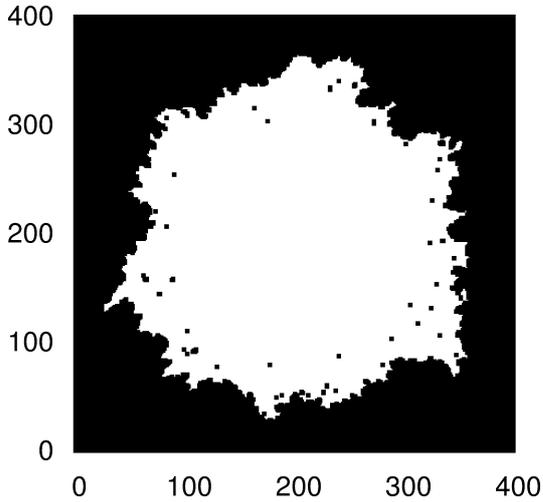}
   \caption{A snap shot of the patch of broken fibers when the load redistribution is among nearest 
surviving neighbours (Model I).
The fiber failure thresholds are uniformly distributed between [$0:1$].}
\label{config1}
\end{figure}
However, the situation can be quite different if one  makes the initial applied load localized (at an 
arbitrarily chosen central site; see Fig. \ref{setup})
in a local load sharing fiber bundle model in greater than one dimension (for one dimension 
the damage interface cannot increase).
Let this load be increased at a slow but constant rate.
Initially no load is present on any fiber except for the one at the central site. As the applied load increases beyond the 
failure threshold of this central
fiber, it breaks and the load carried by it is redistributed among its nearest neighbours and so
on. Here we study two versions of the model. In Model I: In general, whenever a fiber breaks, the load carried by that fiber is 
redistributed equally among its nearest surviving
neighbour(s).  In this way, the fibers which are 
newly exposed to the load, say, after an avalanche, have a relatively low load
compared to the ones which are accumulating load shares from the earlier failures and are still surviving. 
As we shall see later, this helps in maintaining
a compact structure of the cluster or patch of the broken fibers. 
This local force redistribution is justified from the point of view that the newly 
exposed fibers are presumably further away from the point of loading and therefore have to carry 
a smaller fraction of the load at the original central site.

The  fibers on the perimeter of the failed or damaged region,
 which together are carrying the entire load, increase in number with time. Hence the
load per fiber decreases during an ongoing avalanche, which is assumed to be a much faster process compared to  the external
load increase. 
However, as the load 
on the bundle increases at a constant (but slow) rate, the load per fiber along the boundary will tend to increase. 
Eventually, a dynamically stable state will occur when the load per fiber  will
fluctuate around a stable value and the system has reached a self-organized state. 
In this dynamical state,  failure of fibers  in 
the process of avalanches is seen to have a scale free size distribution, suggesting the state to be a 
self-organized critical one. 
We study this model numerically to estimate the avalanche size and other exponent values. 
These values are close to those found for
stochastic sandpile model or Manna model, within our numerical accuracies.

We then study a simpler version of this model, Model II, in which the load of a broken fiber is equally shared by all the
surviving fibers that have atleast one broken neighbour, i.e. it is equally distributed along the boundary of the
broken or damaged patch. Due to the fact that local fluctuations are ignored in the process of load redistribution, 
this version is analytically tractable using a mean-field like approach and the numerical results compare well. 

As can be seen from Fig. \ref{config1}, this version of the fiber bundle model ensures an advancing
interfacial (mode-I) fracture. This topic is widely studied over decades both theoretically
and experimentally (see \cite{bouchaud} for a recent review). Particularly, in the Plexiglas 
experiment \cite{delaplace} two plates were taken 
and disorder was introduced by sandblasting and then were joined together, making a transparent block 
with an easy plane. Interfacial (mode-I)
fracture was then studied with it. A similar situation with this model would arise if one of the 
plates (or softer surface) could be pulled from the middle. The advantage of the present situation for
experiment and in particular for simulations is that one is free from the need of introducing a cut-off
scale by hand as dissipation comes naturally in this model with the increase of effective system size with dynamics. 
This is a more desireable situation since artificial dissipation scales often cause problems in estimating 
the exponent values, which had to avoided by measuring a different quantity instead \cite{lasse}.

Below we first present the numerical results for the nearest neighbour load sharing model (Model I) and then go 
over to the simpler version of 
uniform load sharing along boundary (Model II), 
giving the analytical estimates for the later and comparing them with numerical results.

\section{Model I: Local load and nearest neighbour load redistribution}
\noindent As mentioned before, usually one does not find a finite critical load for local load sharing fiber bundle models.
This is the result of extreme value of statistics, in which the strength of the sample is determined solely by its weakest 
point.
In that case, as the load, no matter how small,  is applied system wide, there will always be a 
large enough weak patch, which will 
be broken and due to
concentration of the load on the boundary and the fact that more load is nucleated if the fracture progresses, 
the patch keeps growing, leading to a system wide failure without any further input from outside. 

However, the situation will be very
different when the initial load is applied locally and not on all elements of the system. 
Initially we apply a load on one fiber, say the one at the middle. We then continue to increase the 
load at a very slow rate. 
When this fiber breaks, the load carried by it is redistributed equally among its four neighbours. 
Since we keep on increasing the load (only on the fibers that are already carrying a non-zero load), this breaking
and redistribution dynamics continues. In general, whenever a fiber breaks, 
the load carried by that fiber is equally shared by the 
surviving fiber(s)
nearest to it.  When the islands break (all four neighbours already broken), 
its load is redistributed among its nearest surviving neighbor(s)
(searched along the perpendicular axes; other searchings are slow and do not change the critical behavior), 
no matter how far they are located. Note that in this way of redistribution, the fibers that are newly exposed to the load
have a rather low load per fiber compared to the ones which are gathering and withstanding the loads of broken fibers from a few
steps earlier. This ensures that the compact structure of the patch of the broken fibers is maintained, since a rather weak but
new fiber is more likely to survive than a stronger but old one, 
thereby eliminating the possibilities of fingering like effects from
the patch boundary. Also note that, following the usual definition of avalanche, 
we do not increase the external load when a fiber breaks or more fibers break due to load redistribution. In other words, the
process of load redistribution happens in a much faster time scale than load increase, which is quasistatic.

\begin{figure}[tb]
\centering \includegraphics[width=9cm]{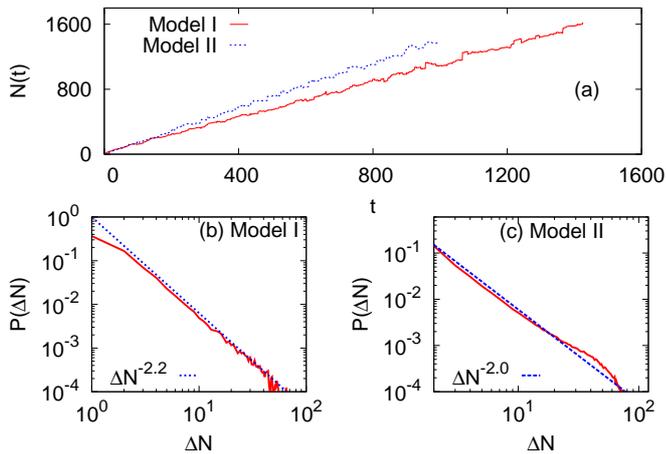}
   \caption{(Color online) (a) The effective system size $N(t)$, i.e. the number of fibers sharing 
the load is plotted against time for both models for a single realisation. It increases on average 
but has jumps of increase and
decrease as well. (b) and (c) show the probability distribution $P(\Delta N)$ of the sizes of 
the jumps $\Delta N=|N(t)-N(t+1)|$ for Model I and Model II respectively, all measured after the time the 
system has reached 
a steady state.}
\label{jump}
\end{figure}
\begin{figure}[tb]
\centering \includegraphics[width=9cm]{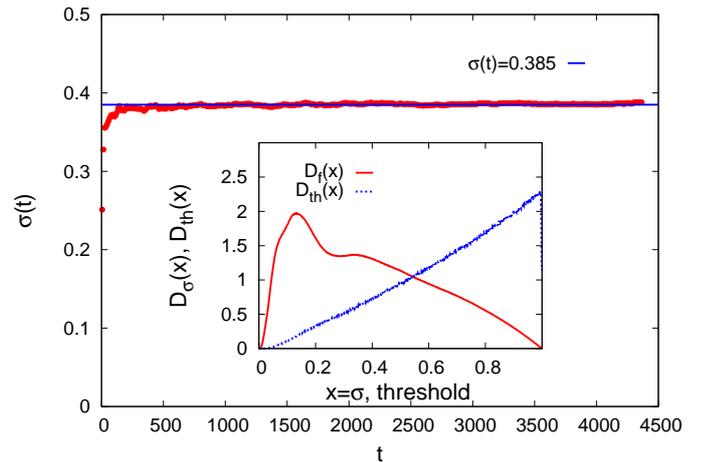}
   \caption{(Color online) The time evolution of the average load per fiber on the broken patch boundary
(the breaking thresholds are distributed uniformly in [$0:1$]) for nearest surviving 
neighbour load redistribution (Model I). 
The saturation is close to $0.385$. The inset shows the stationary distribution of the load per fiber value ($D_f(x)$)
and the failure threshold values ($D_{th}(x)$) of the surviving fibers on the boundary.}
\label{load-sat-loc}
\end{figure}
\subsection{Stationarity in macroscopic quantities}
\noindent As we shall see, the application of load at a point and subsequent localized redistribution rules, bring
the system into a self-organized critical state. One of the signatures of self-organization is the stationarity in 
macroscopic quantities. In this case, the most important quantity is the average value of the load per fiber
\begin{equation}
\sigma(t)=\frac{W(t)}{N(t)},
\end{equation}
where $W(t)$ is the slowly increasing load and $N(t)$ is the effective system size, i.e. number of fibers
carrying the load, which also increases with time (see Fig. \ref{jump}).
We find
that this quantity stabilises, i.e. becomes time independent  after some initial transients. 
The load carried by each fiber is not same. We measured the probability density function $D_f(x)$ of the
load of the fiber, i.e. the probability that a fiber has load between $x$ and $x+dx$ is $D_f(x)dx$.
In fact the entire distribution  of the load per
fiber, and not only its first moment, becomes stable. Similarly, we also find stability in the probability density
 function of the 
failure threshold values of the surviving fibers $D_{th}(x)$, 
each of which carry a non-zero load. Again, $D_{th}(x)dx$ denotes the probability that a 
surviving fiber carrying a non-zero load has the failure threshold value between $x$ and $x+dx$. 
In Fig. \ref{load-sat-loc}  
the saturation of the 
average load per fiber value is shown. The inset shows the stationary distribution of the load per fiber values ($D_f(x)$)
and failure threshold
values ($D_{th}(x)$).  While these results are for the initially uniform (in [$0:1$]) 
threshold distribution, the same phenomenon
occurs for other threshold distributions, like Weibull, Gaussian, triangular etc. The average stationary values 
change for different distributions. Later we will give  analytical estimations of these stationary distributions
for a simplified model, which match well with simulation results.

Note that usually the two basic ingredients of a self-organized critical system are external drive and dissipation
\cite{btw,manna}. In the present case, although we have an external drive, there is no explicit dissipation. But
the effect of dissipation enters the system from the fact that the effective system size ($N(t)$: number of surviving 
fibers carrying non-zero load) is an increasing function of time.

\subsection{Avalanche size distribution}
\begin{figure}[tb]
\centering \includegraphics[width=9cm]{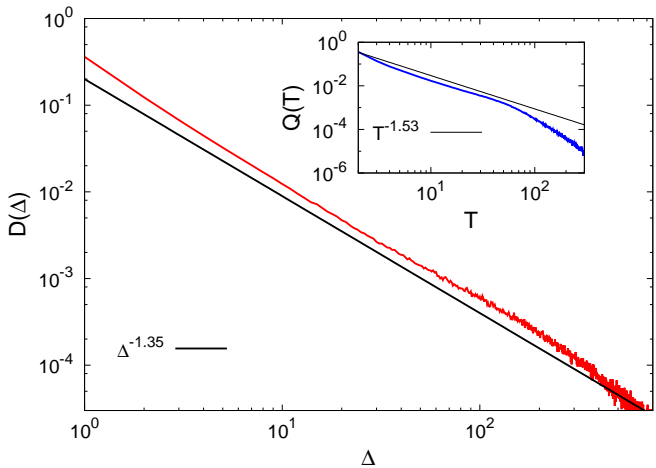}
   \caption{(Color online) The distribution of the avalanche sizes for Model I 
(load redistribution is among nearest surviving neighbours). 
The distribution is a power-law with exponent value $1.35\pm 0.03$. Inset: Distribution of the
duration of an avalanche is shown, which is also a power-law decay with exponent value $1.53\pm 0.02$.}
\label{avalanche-loc-comb}
\end{figure}

\noindent As indicated before, we follow a quasistatic load increase protocol. In effect, this means increasing
the load per fiber uniformly on all those surviving fibers that already carry a non-zero load, until the weakest (having the
smallest difference between load and failure threshold) 
fiber breaks. Then we wait for the system to adjust by breaking the fibers
and redistribution of the loads, and come to a state when no further fiber breaks. All the fibers broken in
between, constitute one avalanche. This process is repeated in every time step.
 The size distribution of avalanches $D(\Delta)$ ($\Delta$ being the size of avalanche, i.e. 
number of fibers broken in an avalanche) measured in this way are plotted
in Fig. \ref{avalanche-loc-comb}. This shows a power-law distribution $D(\Delta)\sim \Delta^{-\gamma}$ 
with exponent value close to $1.35\pm 0.03$. 
  This exponent value remains unchanged when other types of threshold
distributions (not shown here) are used. We have also measured the duration of an avalanche. 
This is defined as the number of times ($T$) 
the load is redistributed during one process of avalanche. The probability distribution
of duration $Q(T)$  is also a power-law decay ($Q(T)\sim T^{-\omega}$) with exponent value close
to $\omega=1.53\pm 0.02$ (see inset of Fig. \ref{avalanche-loc-comb}).
It is interesting to note that avalanche size distribution exponent 
 value is rather close to the stochastic sandpile or Manna universality class \cite{manna,lubeck}.

Apart from these, the size distributions of the changes in effective system size ($\Delta N=|N(t)-N(t+1)|$)
was measured, again after the system has reached a stationary state. This distribution is also a power-law
with exponent value close to $2.20\pm 0.03$ (see Fig. \ref{jump}).

\section{Model II: Local load and uniform load redistribution along broken patch boundary}
\begin{figure}[tbh]
 \includegraphics[width=7cm]{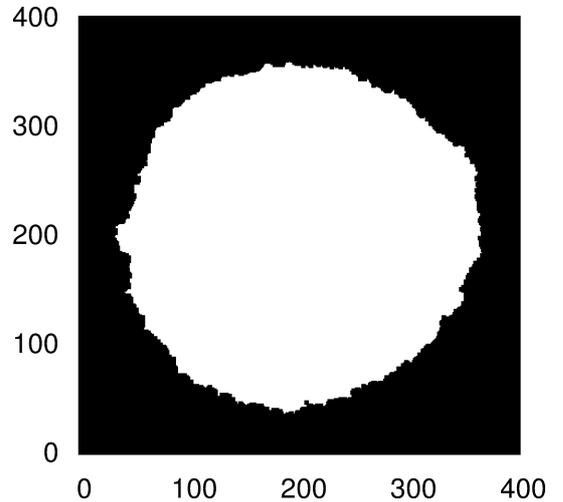}
   \caption{A snap shot of the patch of broken fibers when the load redistribution is uniform 
along the broken patch boundary, i.e., Model II.
The failure thresholds are uniformly distributed between [$0:1$].}
\label{config2}
\end{figure}

\noindent A simpler version of the model studied so far is the one where the load carried by a broken fiber
is equally redistributed among all the fibers that have at least one broken neighbour. This is another way
of saying that the load is redistributed uniformly along the entire  boundary, once a fiber breaks (see Fig. \ref{config2}).  
Since the size of the boundary is large after the transients, the behaviour of the 
model in the present form is tractable via a mean field like approach. Below we first discuss the
stationary distribution functions of load per fiber value and failure thresholds for different
initial failure threshold distributions; comparing the estimates with numerical results. Then we
do the same for avalanche size distribution.   
\begin{figure}[tb]
\centering \includegraphics[width=9cm]{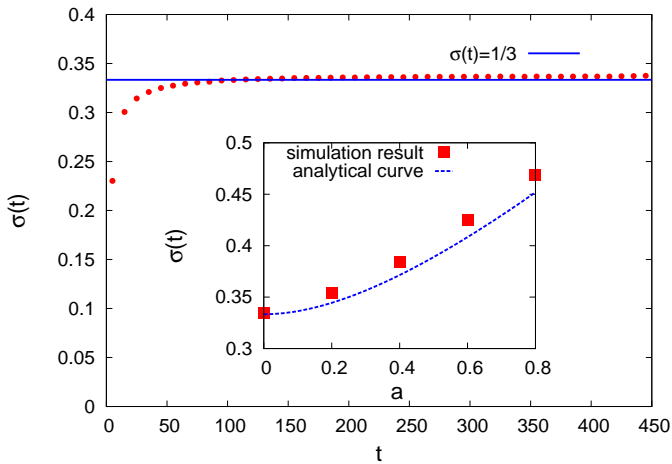}
   \caption{(Color online) The time evolution of the average load per fiber  
(when the threshold is uniformly distributed between [$0:1$])
for Model II (i.e., uniform load redistribution along the patch boundary). 
The solid line is the analytical estimate ($1/3$) and the simulation 
result is close to that value. The inset shows the comparison of the simulation results and analytical estimates of the saturation
load per fiber value, when the threshold is distributed uniformly between [$a:1$].}
\label{load-sat}
\end{figure}
\begin{figure}[tb]
\centering \includegraphics[width=9cm]{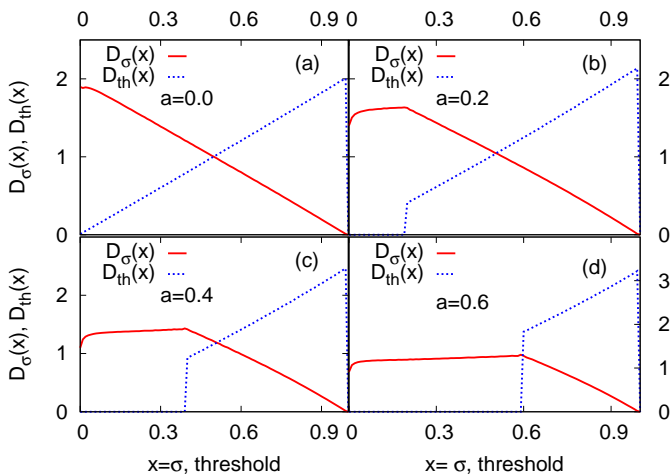}
   \caption{(Color online) Simulation results for the probability density functions  of the load per 
fiber (red continuous lines) and failure thresholds of the fibers (blue dotted lines),
both only along the boundary of the patch of broken fibers, for different values of the lower
cut-offs in the otherwise uniform threshold distributions for Model II (uniform load redistribution along
the patch boundary).}
\label{dist}
\end{figure}
\subsection{Stationary states}
\noindent Same as the earlier version of the model, the load distribution and the threshold
distribution along the patch boundary reaches a stationary state after some transient. 
In this case, however, one can also
make an analytical estimate along with the numerical simulations, to get a good agreement.
\subsubsection{Initial threshold distribution is uniform}
\noindent Consider the failure thresholds to be distributed uniformly between the limits [$a:1$]. 
First let us take the case when $a=0$. We follow the algorithm for load increase as stated before. If one measures
the average load per fiber with time $\sigma(t)$ (one time step being one avalanche), one finds that after a transient 
it saturates to a given value
(see Fig. \ref{load-sat}). For this case, the value is close to $1/3$.  In analytically estimating it
 we assume that  dynamically all values of force 
(between [$0:1$], i.e. the range of the threshold distribution)
are generated with equal probability.
So, the probability that a load between $x$ and $x+dx$ is applied on the fibers is $c_1dx$ (unnormalized; $c_1$
is a constant independent of $x$ since the force distribution is assumed to be uniform). Now among these fibers,
only the ones with failure threshold greater than $x$ will survive. The probability that the fibers have 
failure threshold greater than $x$ is proportional to $(1-x)$ (since we are considering threshold
distribution to be unuiform in [$0:1$]). Therefore, the probability that a surviving fiber has a load
between $x$ and $x+dx$ is the joint probability that a load between $x$ and $x+dx$ is applied on the fibers
{\it and} the fibers have failure thresholds higher than $x$, which is given by $c_1(1-x)dx$. From
normalization $c_1\int\limits_0^1(1-x)dx=1$, one has $c_1=2$. So, the probability density function of
the surviving fibers having load  $x$ is $D_{\sigma}(x)=2(1-x)$, which compares well
with numerical estimate (Fig. \ref{dist}). Of course, the average load per fiber is 
$\int\limits_0^1D_{\sigma}(x)=\int\limits_0^12(1-x)xdx=\frac{1}{3}$, which is again close to what 
we see numerically (Fig. \ref{load-sat}). Similarly, the probability that the fibers in the system have
a failure threshold between $x$ and $x+dx$ is $c_2dx$ (unnormalized; $c_2$ is independent of $x$ since the
threshold distribution is uniform in [$0:1$]). Now, the probability that a load equal to or less than
$x$ is applied on the fibers is proportional to $x$ (since force distribution is uniform). Hence the 
probability that a surviving fiber has a threshold between $x$ and $x+dx$ is the joint probability of
a fiber having its threshold between $x$ and $x+dx$ {\it and} load below $x$. This probability
is $c_2xdx$. Normalization gives $\int\limits_0^1c_2xdx=1$ i.e., $c_2=2$. So, the probability density function
for the threshold to be $x$ is $D_{th}(x)=2x$. This also compares well with simulations in Fig. \ref{dist}.

Now  we go to the case when $a>0$, i.e., the threshold distribution function has a lower cut-off. 
Now, since all fibers have
threshold higher than $a$ and we have assumed before that force is uniformly distributed, the probability density
function for force will be uniform between [$0:a$] and then decreasing linearly as before (can be obtained along the same 
line as before). From normalization condition, the
height of the uniform part is $\frac{2}{a+1}$. Therefore, one gets
\begin{eqnarray}
D_{\sigma}(x) &=& \frac{2}{a+1} \quad \mbox{for} \quad x<a \nonumber \\
&=& \frac{2}{a+1}\left[1-\frac{x-a}{1-a}\right]\quad \mbox{otherwise}
\end{eqnarray}
This is what we get numerically as well (see Fig. \ref{dist})
for different values of $a$ (for $a=0$ we get back $2(1-x)$). In the same line one can also 
calculate the probability density
function for threshold with finite $a$ 
\begin{eqnarray}
D_{th}(x)&=& 0 \quad \mbox{for} \quad x<a \nonumber \\
&=& \frac{2x}{1-a^2} \quad \mbox{otherwise.}
\end{eqnarray}
This compares well with numerical result (also shown in Fig. \ref{dist}). 
If we now calculate the average force, it comes out to be 
\begin{equation}
\sigma(t\to\infty)=\int\limits_0^1xD_{\sigma}(x)dx=\frac{a^2}{1+a}+\frac{2}{1-a^2}\left[\frac{1}{6}-\frac{a^2}{2}+\frac{a^3}{3}\right],
\label{avgsigma}
\end{equation}   
We compare the prediction of this calculation with the values 
obtained numerically
in Fig. \ref{load-sat}. Of course, we get back $1/3$ for $a=0$.
\subsubsection{Initial threshold distribution is Weibull}
\noindent To show universality of the above phenomenon with respect to different threshold distributions, we have 
checked it for different threshold distributions (Gaussian, triangular etc.). Here we show the case of Weibull
distribution, which is more abundant in real situations.

\begin{figure}[tb]
\centering \includegraphics[width=9cm]{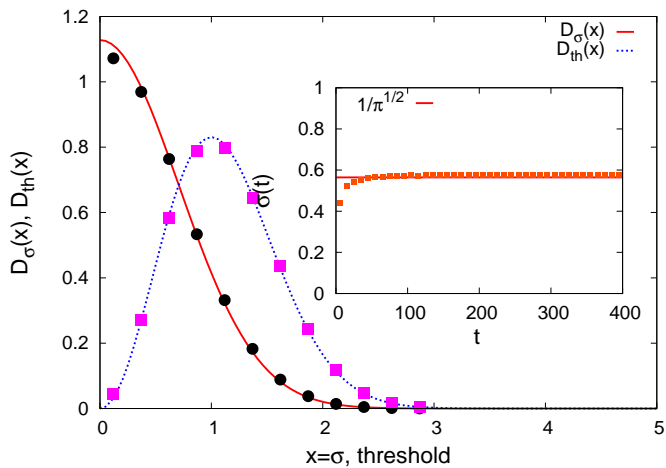}
   \caption{(Color online) The probability density functions of load per fiber ($D_f(x)$) and failure thresholds ($D_{th}(x)$)
of the fibers in the boundary of the broken patch for Weibull threshold distribution with $\alpha=2$ 
and $\beta=1$ for Model II. The solid lines are analytical 
predictions (Eqs. (\ref{wei-force}) and (\ref{wei-th})) 
and the points are simulation results. The inset shows the time variation of the average load per 
fiber value, which saturates close to the analytical estimate $1/\sqrt{\pi}$.}
\label{weibull}
\end{figure}
A general form of the Weibull distribution is 
\begin{equation}
W_{\alpha,\beta}(x)=\alpha \beta x^{\alpha-1}e^{-\beta x^{\alpha}},
\end{equation}
 where $\alpha$ and $\beta$
are two parameters. Let us consider the particular case when $\alpha=2$ and $\beta=1$. As before, the probability
that the threshold is greater than $x$ is proportional to $\int\limits_{x}^{\infty}x^{\prime}e^{-{x^{\prime}}^2}dx^{\prime}\sim e^{-x^2}$.
Now since the probability density function for force is uniform, the probability of a fiber having load
between $x$ and $x+dx$ is $e^{-x^2}P(x)dx$, with $P(x)=c$ (unnormalized) since the force distribution is assumed 
to be uniform. As before, the normalization condition would require $c\int\limits_0^{\infty}e^{-x^2}dx=1$ 
giving $c=\frac{2}{\sqrt{\pi}}$. Hence the normalized probability density function for the load on the surviving fibers
will be
\begin{equation}
D_{\sigma}(x)=\frac{2}{\sqrt{\pi}}e^{-x^2}.
\label{wei-force}
\end{equation}
 Similarly, the probability that the load is lower than $x$
is proportional to $x$. Using the form for threshold distribution ($\sim xe^{-x^2}$),
 it is straightforward to get the probability density
function for threshold distribution of the survived fibers, which is 
\begin{equation}
D_{th}(x)=\frac{4}{\sqrt{\pi}}x^2e^{-x^2}.
\label{wei-th}
\end{equation}
 Both
of these functions are in good agreement with numerical simulations (see Fig. \ref{weibull}). 
Also the saturation value of the average load per fiber
can be calculated  as
\begin{equation}
 \int\limits_0^{\infty}xD_{\sigma}(x)dx=\frac{2}{\sqrt{\pi}}\int\limits_0^{\infty}xe^{-x^2}dx=\frac{1}{\sqrt{\pi}},
\end{equation}
which is also in good agreement with simulation value (see inset Fig. \ref{weibull}).

\subsection{Avalanche size distributions}
\begin{figure}[tb]
\centering \includegraphics[width=9cm]{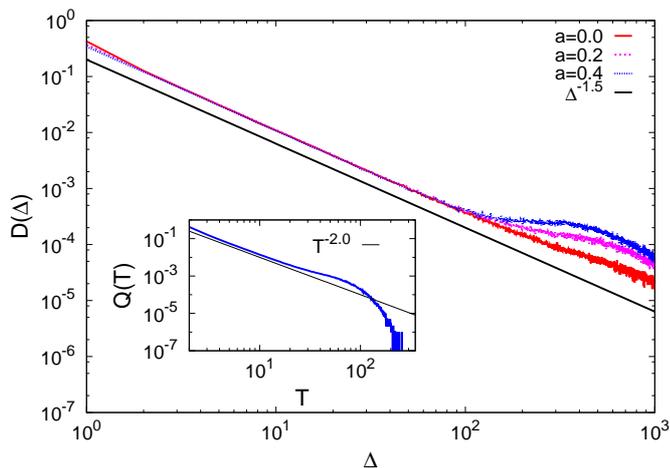}
   \caption{(Color online) The distribution of the avalanche sizes are plotted for zero and finite lower cut-offs
for Model II. 
The distribution function is a power-law with exponent value $1.50\pm 0.01$, which
is also our estimate from scaling arguments. Inset: The distribution of the duration of an avalanche is plotted for Model II. 
This shows a power-law decay
with exponent value $2.00\pm 0.01$.}
\label{avalanche-comb}
\end{figure}

\noindent The size distribution of avalanches are measured as before. Now we find the size distribution exponent to be
close to $3/2$ (see Fig. \ref{avalanche-comb}), which is in agreement with the scaling prediction of avalanche 
size distributions in SOC models for mean field
\cite{lubeck}, i.e. $\gamma=1+\frac{\delta}{1+\theta+\delta}$, where $\delta$ and $\theta$ are the surviving 
probability decay exponent and number of particle growth exponent respectively, both of which
are 1 in mean field, giving $\gamma=3/2$. Note that similar exponent value was
obtained before when one measures avalanches only near the critical point in global load sharing 
models \cite{pradhan5} and also
in other versions of self-organized models with fiber regeneration \cite{moreno}.

We also measure the distribution of avalanche duration. The duration distribution is a 
power-law with exponent value close to $2.00\pm 0.01$ (see inset of Fig. \ref{avalanche-comb}), 
which is again in agreement with
scaling  predictions of SOC models in mean field, i.e. $\omega=1+\delta$. 
The size distribution of the changes in effective system size is 
also close to $2.00\pm 0.05$ (see Fig. \ref{jump}).

Now to estimate the value of the avalanche size exponent, we first assume that the average load per fiber on the 
perimeter of the
damaged region has a 
distribution which is Gaussian around its mean: $P(\sigma )\sim e^{-(\sigma-\sigma_c)^2/{\delta \sigma}}$. Hence, on a
dimensional analysis,  mean-squared
fluctuation $\delta \sigma\sim (\sigma-\sigma_c)^2$ .
Also the avalanche size $\Delta$ scales as $(\delta \sigma)^{-1}$ since it may be viewed as the number of broken fibers
after a load increase of $\delta \sigma$. This gives
\begin{equation}
(\sigma-\sigma_c)\sim \Delta^{-1/2}.
\label{sig1}
\end{equation} 
The probability of an avalanche being of the size between
$\Delta$ and $\Delta+d\Delta$ is $D(\Delta)d\Delta$. Now, the deviation from the critical point scales \cite{pradhan2.5}
with the cumulative size of all avalanches upto that point;
giving $(\sigma-\sigma_c)\sim \int\limits_{\Delta}^{\infty}D(\Delta)d\Delta$. If we take $D(\Delta)\sim \Delta^{-\gamma}$,
then 
\begin{equation}
(\sigma-\sigma_c)\sim \Delta^{1-\gamma}.
\label{sig2}
\end{equation}  
Comparing Eqs. (\ref{sig1}) and (\ref{sig2}) therefore we have $\gamma=\frac{3}{2}$. So the probability density function for
the avalanche size becomes $D(\Delta)\sim \Delta^{-3/2}$, which fits well with simulation results (Fig. \ref{avalanche-comb}).

Note that we have also checked these results for triangular lattices, where the nearest neighbours are six. 
As one can see, the analytical estimates are independent of lattice topology, hence predict same behaviour 
for triangular lattice as well, which is 
what we get numerically as well (not shown here).

\section{Summary and conclusion}
\noindent Here we study the local load sharing fiber bundle model in a square lattice, with the 
load being initially applied only at a central site. The load is slowly increased and a broken
patch grows in the middle of the system, the boundary of which carries the entire load. Since
the boundary increases, the load per fiber reaches a stationary value after some time and 
responses of the system in that state is scale free, indicating a self-organized critical
state for the system.


We study two versions of this model. In Model I: After the failure of the central fiber, the load is shared 
by its four nearest neighbouring sites, which initially had no load on them. Once any of them fails, its load will be shared
equally by its nearest surviving fiber(s)  and so on.
These load shares accumulate on all the fibers on the perimeter and grows with time since their
respective appearance on the perimeter of the damaged central region. 
This may be interpreted as the fact that further the surviving fibers 
are from the
central point, lower are the loads on them and closest 
fiber (also the longest surviving fiber on the perimeter) to the central site
accumulates the maximum load share.      
As stated before, the model reaches a stationary state after some transients.
This leads to a self-organized dynamics of failure in the model. We have studied the breaking statistics in such a 
self-organized dynamical state.  

 While the usual ingredients for
self organization are external drive and dissipation \cite{btw,manna}, in our model the latter 
is apparently absent. However a similar  effect 
is coming from the fact that the effective system size, i.e. the perimeter length of the patch of the broken fibers, is not
a constant but increases with the increase  of load with time. 
This leads the system into a self-organized critical state, from which if one makes 
a departure (by increasing the load) the response of the system 
(size of the avalanche and its duration) can be of any size, leading to
scale free dynamics. Particularly, we observe, in the avalanche size distribution, exponent value 
is $1.35\pm 0.03$ (see Fig. \ref{avalanche-loc-comb}) 
and the duration distribution of 
an avalanches has an exponent value $1.53\pm 0.02$ (see inset of Fig. \ref{avalanche-loc-comb}).
Also the change in the effective system size follows a power-law with exponent close to $2.2$ (see Fig. \ref{jump}). 
We have checked, the
exponent values do not change if the identities of the nearest neighbour fibers 
(along lattice axes, or along diagonals as well)
are changed. Also we have checked the universality using different (e.g., Weibull, uniform between [$a:1$] and so on)
 fiber strength distributions.
These exponent values are close to what are seen in Manna model. 

We then  studied a simpler version of this model, Model II, where the load redistribution is uniform along the boundary
of the broken patch. This reduces the fluctuation in the system significantly (see Figs. \ref{config1},\ref{config2} 
for a visual comparison),
and brings the system to a mean-field limit. The average load per fiber value still saturates to a finite value.
 We have made an estimate of this average value for a system with uniform threshold distribution between [$a:1$], and also for
a Weibull distribution with certain values of parameters.
 It turns out, for uniform distribution with $a=0$, the probability density function for a
surviving fiber to have load $x$ can be calculated to be $2(1-x)$  and 
 average value of $\sigma(t\to \infty)$ is estimated to be $1/3$,
which is in close agreement with simulation results. However, for non-zero values of $a$, 
we find significant departure from the analytical estimate
 (given by Eq. (\ref{avgsigma})). Similar estimates for Weibull distribution with $\alpha=2$ and $\beta=1$ is made,
 where this saturation
value comes out to be $1/\sqrt{\pi}$, again very close to simulation results. 

We also present a scaling theory to predict the avalanche size distribution in this case. The exponent value
 for the size distribution 
is estimated to be $3/2$, which is in close agreement with the numerical results (Fig. \ref{avalanche-comb}). 
This is of course
the mean field limit of self-organized criticality \cite{lubeck}. The exponent value ($2.00\pm 0.02$; 
see inset of  Fig. \ref{avalanche-comb})
for the distribution of duration of avalanche also matches with the
mean field estimate.

Note that one could not have obtained this self-organization in an one dimensional array of local load sharing
fibers,  due to the finite extension of the damaged region
(two fibers on the two sides of the damaged region will have to carry the load). 
This self-organization  therefore occurs only for two and 
higher dimensions. Also note that a long range load transfer,
with initial load along a line, requires an explicit dissipation term to get self-organized dynamics \cite{petri}. 
Similar efforts were made in capturing the morphology of the fractured surfaces by considering the elastic modulus
of the lower plane of the model \cite{hansen}. 

In  conclusion, the self-organizing dynamics of failure in this local load sharing fiber bundle model 
(in more than one dimension) has interesting
critical (scale free) behaviour. The models show mode-I damage propagation without having
to put a dissipation scale by hand. In the nearest neighbour load sharing version (Model I), 
the avalanche size distribution follows a power-law with exponent value close to $1.35\pm 0.03$. 
The distribution for the duration of avalanche follows a power-law with exponent value close to $1.53\pm 0.02$.
 We have checked that these
responses remain unchanged with different types of distribution of the breaking thresholds of the 
fibers and both square and triangular lattices.
We have also studied a simpler version of this model, where the load redistribution is uniform along broken 
patch boundary (Model II), 
where mean-field approximations work and 
find good agreement between analytical
predictions and numerical simulations.
\acknowledgements
\noindent  The authors acknowledge the
participants of the INDNOR Project meeting, particularly Alex Hansen, Knut  M\aa l\o y, 
Srutarshi Pradhan and Purusattam Ray, 
for useful discussions. The 
financial supports from project INDNOR 217413/E20 (NFR, Govt. of Norway) and BKC's JC Bose Fellowship (DST, Govt. of India) are 
acknowledged.

\end{document}